\documentclass{aa}

\usepackage{graphicx}
\usepackage{txfonts}
\DeclareRobustCommand{\VAN}[3]{#2}
\let\VANthebibliography\thebibliography
\def\thebibliography{\DeclareRobustCommand{\VAN}[3]{##2}\VANthebibliography}

\usepackage{color}

\usepackage{textcomp}
\usepackage{amsmath}    
\usepackage{amssymb}    
\begin{document}

   \title{Origin of the ring structures in Hercules A}
   \subtitle{Sub-arcsecond 144\,MHz to 7\,GHz observations\thanks{The reduced images are only available at the CDS via anonymous ftp to cdsarc.u-strasbg.fr (130.79.128.5) or via http://cdsarc.u-strasbg.fr/viz-bin/qcat?J/A+A/XXX/YYY}}
   \titlerunning{Origin of the ring structures in Hercules A;  sub-arcsecond 144\,MHz to 7\,GHz observations}

   \author{R. Timmerman\inst{1}\fnmsep\thanks{E-mail: rtimmerman@strw.leidenuniv.nl (RT)}
        \and
        R. J. van Weeren\inst{1}
        \and
        J. R. Callingham\inst{1,2}
        \and
        W. D. Cotton\inst{3}
        \and
        R. Perley\inst{4}
        \and
        L. K. Morabito\inst{5,6}
        \and
        N. A. B. Gizani\inst{7} 
        \and
        A. H. Bridle\inst{3}
        \and
        C. P. O'Dea\inst{8}
        \and
        S. A. Baum\inst{8}
        \and
        G. R. Tremblay\inst{9}
        \and
        P. Kharb\inst{10}
        \and
        N. E. Kassim\inst{11}
        \and
        H. J. A. Röttgering\inst{1}
        \and
        A. Botteon\inst{1}
        \and
        F. Sweijen\inst{1}
        \and
        C. Tasse\inst{12,13}
        \and
        M. Brüggen\inst{14}
        \and
        J. Moldon\inst{15}
        \and
        T. Shimwell\inst{1,2}
        \and
        G. Brunetti\inst{16}
    }

   \institute{Leiden Observatory, Leiden University, P.O. Box 9513, 2300 RA Leiden, The Netherlands
        \and
        ASTRON, Netherlands Institute for Radio Astronomy, Oude Hoogeveensedijk 4, Dwingeloo, 7991 PD, The Netherlands
        \and
        National Radio Astronomy Obs, 520 Edgemont Rd., Charlottesville, VA 22903, USA
        \and
        National Radio Astronomy Obs, P.O. Box 0, Socorro, NM 87801, USA
        \and
        Centre for Extragalactic Astronomy, Department of Physics, Durham University, Durham DH1 3LE, UK
        \and
        Institute for Computational Cosmology, Department of Physics, University of Durham, South Road, Durham DH1 3LE, UK
        \and
        Hellenic Open University, School of Science \& Technology, Parodos Aristotelous 18, Perivola Patron, Greece
        \and
        Department of Physics \& Astronomy, University of Manitoba, Winnipeg, MB R3T 2N2, Canada
        \and
        Harvard-Smithsonian Center for Astrophysics, 60 Garden St., Cambridge, MA 02138, USA
        \and
        National Centre for Radio Astrophysics, S. P. Pune University Campus, Post Bag 3, Ganeshkhind Pune 411 007, India
        \and
        Remote Sensing Division, Naval Research Laboratory, Code 7213, 4555 Overlook Ave SW, Washington DC 20375, USA
        \and
        GEPI \& USN, Observatoire de Paris, CNRS, Universit\'e Paris Diderot, 5 place Jules Janssen, 92190 Meudon, France
        \and
        Centre for Radio Astronomy Techniques and Technologies, Department of Physics and Electronics, Rhodes University, Grahamstown 6140, South Africa
        \and
        Hamburger Sternwarte, University of Hamburg, Gojenbergsweg 112, 21029 Hamburg, Germany
        \and
        Instituto de Astrof\'isica de Andaluc\'ia (IAA, CSIC), Glorieta de las Astronom\'ia, s/n, E-18008 Granada, Spain
        \and
        INAF—Istituto di Radioastronomia, Via Gobetti 101, I-40129 Bologna, Italy
    }

   \date{Received XXX; accepted YYY}

 
  \abstract{The prominent radio source Hercules\,A features complex structures in its radio lobes. Although it is  one of the most comprehensively studied sources in the radio sky, the origin of the ring structures in the Hercules\,A radio lobes remains an open question. We present the first sub-arcsecond angular resolution images at low frequencies ($<$300~MHz) of Hercules\,A, made with the International LOFAR Telescope. With the addition of data from the Karl G. Jansky Very Large Array, we mapped the structure of the lobes from 144~MHz to 7~GHz. We explore the origin of the rings within the lobes of Hercules\,A, and test whether their properties are best described by a shock model, where shock waves are produced by the jet propagating in the radio lobe, or by an inner-lobe model,  where the rings are formed by decelerated jetted plasma. From spectral index mapping our large frequency coverage reveals that the curvature of the different ring spectra increases with distance away from the central active galactic nucleus. We demonstrate that the spectral shape of the rings is consistent with synchrotron aging, which speaks in favor of an inner-lobe model where the rings are formed from the deposition of material from past periods of intermittent core activity.}

   \keywords{large-scale structure of Universe -- galaxies: active -- radio continuum: galaxies -- radiation mechanisms: non-thermal -- galaxies: clusters: individual: Hercules A}

   \maketitle
%

\section{Introduction}

\begin{figure*}[p]
    \centering
    \includegraphics[width=0.88\textwidth]{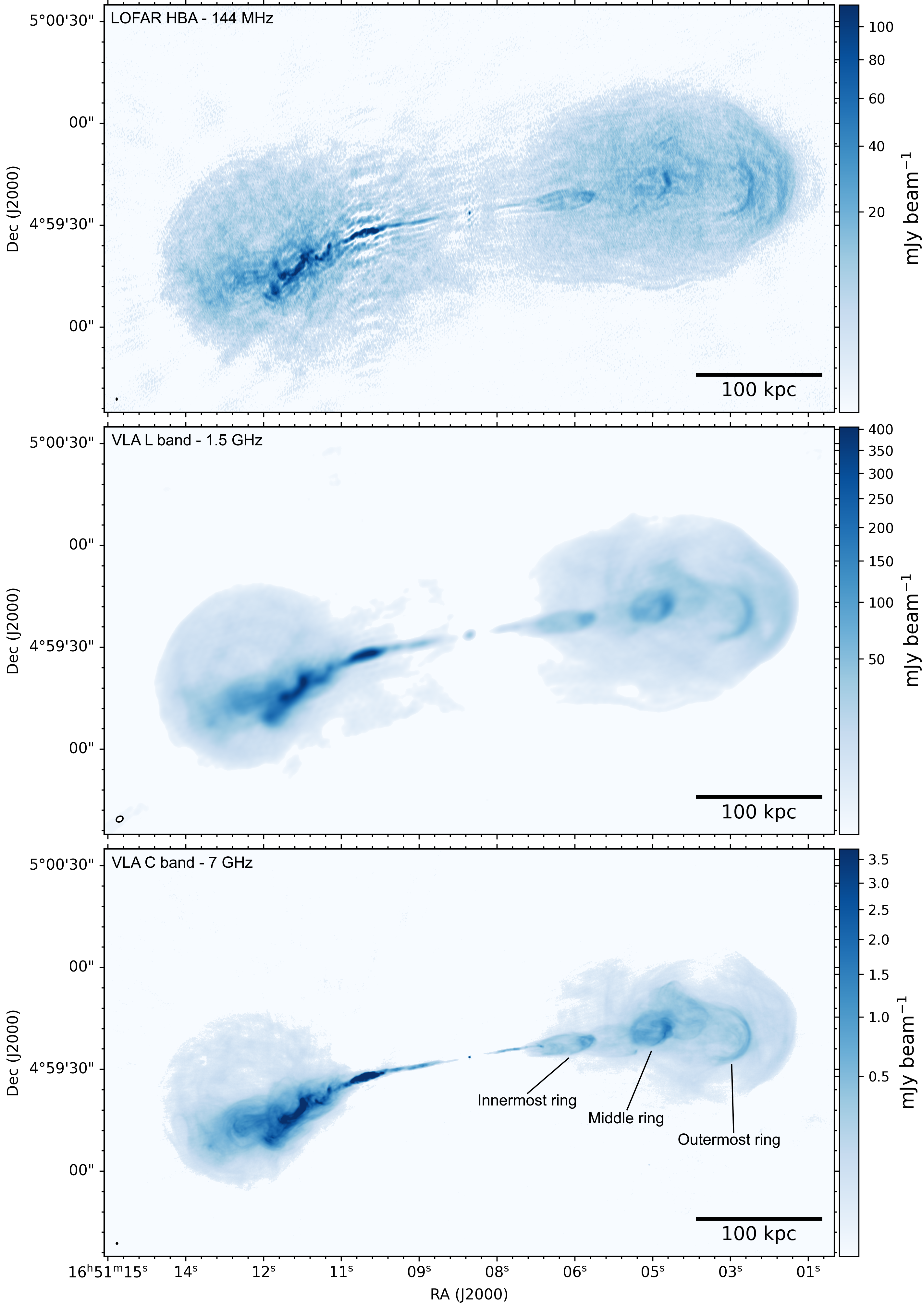}
    \caption{Radio images of Hercules\,A constructed from the LOFAR HBA (144~MHz, top panel), the VLA L band (1.5~GHz, middle panel), and VLA C band (7~GHz, bottom panel). The scale bar in the bottom right corner of each panel measures 100~kpc at the redshift of Hercules\,A. The color scale of each image goes from three times the rms noise level to the peak brightness. The synthesized beam sizes of the three images are indicated by the white oval in the bottom left corner of each panel, and are summarized along with the rms noise levels in Table \ref{tab:img1}.}
    \label{fig:images}
\end{figure*}

The origin of the three bright rings in the lobes of the notable radio source Hercules\,A \citep{bolton48} has remained unclear since their identification by \citet{dreher84}. Understanding the formation of such rings in the radio lobes of active galactic nuclei (AGN) provides insight into the dynamics of the relativistic jets, the role of shock waves, and particle acceleration in the lobes of radio AGN. Hercules\,A is associated with a central dominant (cD) galaxy located in the center of a galaxy cluster at a redshift of \(z=0.154\) \citep{greenstein62}, and is known for its extreme brightness and large angular scale. Hercules\,A spans approximately 190 by 60~arcseconds on the sky (530 by 170~kpc), and has a flux density of roughly 45~Jy at 1.4~GHz, enabling detailed studies of the structure of the entire source, from the freshly emitted jets to the old diffuse plasma in the lobes. Therefore, the source is used as a standard to assess the physics of fainter lobed AGN. 

Hercules\,A displays complex structures in its lobes that have been difficult to interpret relative to the standard Fanaroff-Riley (FR) I and II~schemes \citep{fanaroff74,meier91}. The source has two bright radio lobes; however, the characteristic hotspots of FR II-type radio galaxies are absent. Similarly, while Hercules\,A is jet-dominated, the lobes feature hard outer edges instead of fading away into the intracluster medium (ICM), as is typical for an FR I-type source. For these reasons, Hercules\,A is generally categorized as an intermediate FR I/II~source \citep[e.g.,][]{meier91,sadun02, saxton02,gizani03}. The eastern lobe is dominated by a bright jet from the AGN that slowly diffuses into the lobe, while the western lobe mainly shows a very distinct series of three rings \citep{dreher84, mason88} at projected distances of between 55 kpc and 230 kpc from the host galaxy. 
The jets from the AGN produce Alfvénic perturbations that cross the relativistic plasma of the lobes and induce brightness fluctuations.
Recent investigations by \citet{gizani99, gizani03, gizani04} used both radio observations from the Very Large Array (VLA) and X-ray observations from the Röntgensatellit (ROSAT) to study Hercules\,A. They concluded, based on the observed rotation measure features and an analytical model for the magnetic field strength, that the eastern jet is orientated towards us while the western jet is receding from us at an inclination angle of 50\(^\circ\) relative to the line of sight. In addition, they suggested that the apparent absence of a jet structure in the western lobe could be due to Doppler dimming, which could be as strong as a factor of \(\sim\,20\) \citep{gizani03}. Subsequent research by \citet{gizani05} at 74 MHz and 325 MHz using the VLA--Pie Town link \citep{lane05} found that the spectral differences between the jets and the lobes at these low frequencies are smaller than those observed at higher frequencies \citep{gizani03}. Such spectral differences imply the presence of regions of various ages, thereby supporting the multiple outbursts interpretation. Unfortunately, the limited angular resolution of the VLA--Pie Town link (\(\sim 10\) arcseconds at 74~MHz) did not permit a detailed investigation of the rings.

Despite the many studies focused on understanding the structure of Hercules\,A, a few key features of the lobes remain a mystery. Of particular interest for this paper is the origin of the rings present in the western lobe. Two models have been proposed \citep[e.g.,][]{mason88,meier91,morrison96,saxton02,gizani03}: either the rings are caused by a series of shocks between the old lobe material and new material from jet outbursts or the rings form the surfaces of multiple inner lobes associated with separate AGN outbursts. 

In the shock model, adiabatic compression and particle acceleration are assumed to be responsible for producing thin and curved emission regions, which closely match the observed morphology of the rings. Although it is an idealized assumption that the compression is adiabatic, this is a commonly employed assumption when studying shock models in Hercules\,A and other sources \citep[e.g.,][]{meier91, bruggen07, jubelgas07}. However, the shock model is difficult to match with the spectral indices of the rings, which are relatively flat compared to the lobes, even at low frequencies. This implies that the shock wave induces a significant amount of particle acceleration. Although the observed spectral indices are not unusual for shock waves, it is peculiar that the outermost ring in the western lobe features a significantly steeper spectrum. Assuming that the lobe spectrum evolves mainly due to aging, the spectral index of the outermost ring could be obtained solely through adiabatic compression. In addition, the spectra of the rings have been found to be similar to that of the eastern jet, which would have to be merely a coincidence \citep{gizani03}. 

In the second scenario the rings could be formed as the surface of inner lobes. As the jetted plasma decelerates, its beaming is reduced and its apparent brightness increases. This also provides an explanation for the lack of visible rings in the eastern lobe, where this effect works in the opposite way. Time retardation between the two lobes due to their inclination to the line of sight can also break their symmetry. Without invoking any particle acceleration, the steep spectrum of the outermost ring is no longer problematic. However, this strong rim-brightening is generally not observed in lobes.

Using new International LOFAR Telescope observations at 144~MHz, and Karl G. Jansky Very Large Array (VLA) observations in both the L and C~bands, we aim to study the spectral properties of Hercules\,A between 144~MHz and 7~GHz at sub-arcsecond angular resolution. 
This wide range of observing frequencies provides an unparalleled data set to study the spectral properties of the source, and to test for any spectral turnovers due to absorption mechanisms, which generally occur at frequencies below 1~GHz. Furthermore, the unprecedented angular resolution and sensitivity at low frequencies provide a large lever arm in frequency space, facilitating an investigation into the origin of the rings in the Hercules\,A lobes.

In this paper we adopt a \(\Lambda\)CDM cosmology with a Hubble parameter of \(H_0\ =\ 67.4~\mathrm{km}\ \mathrm{s}^{-1}\ \mathrm{Mpc}^{-1}\), a matter density parameter of \(\Omega_m\ =\ 0.315\), and a dark energy density parameter of \(\Omega_\Lambda\ =\ 0.685\) \citep{planck18}. We define our spectral indices \(\alpha\) according to \(S \propto \nu^\alpha\), where \(S\) is flux density and \(\nu\) is frequency.

\section{Observations and data reduction}

\begin{figure*}[p]
    \centering
    \includegraphics[width=0.92\textwidth]{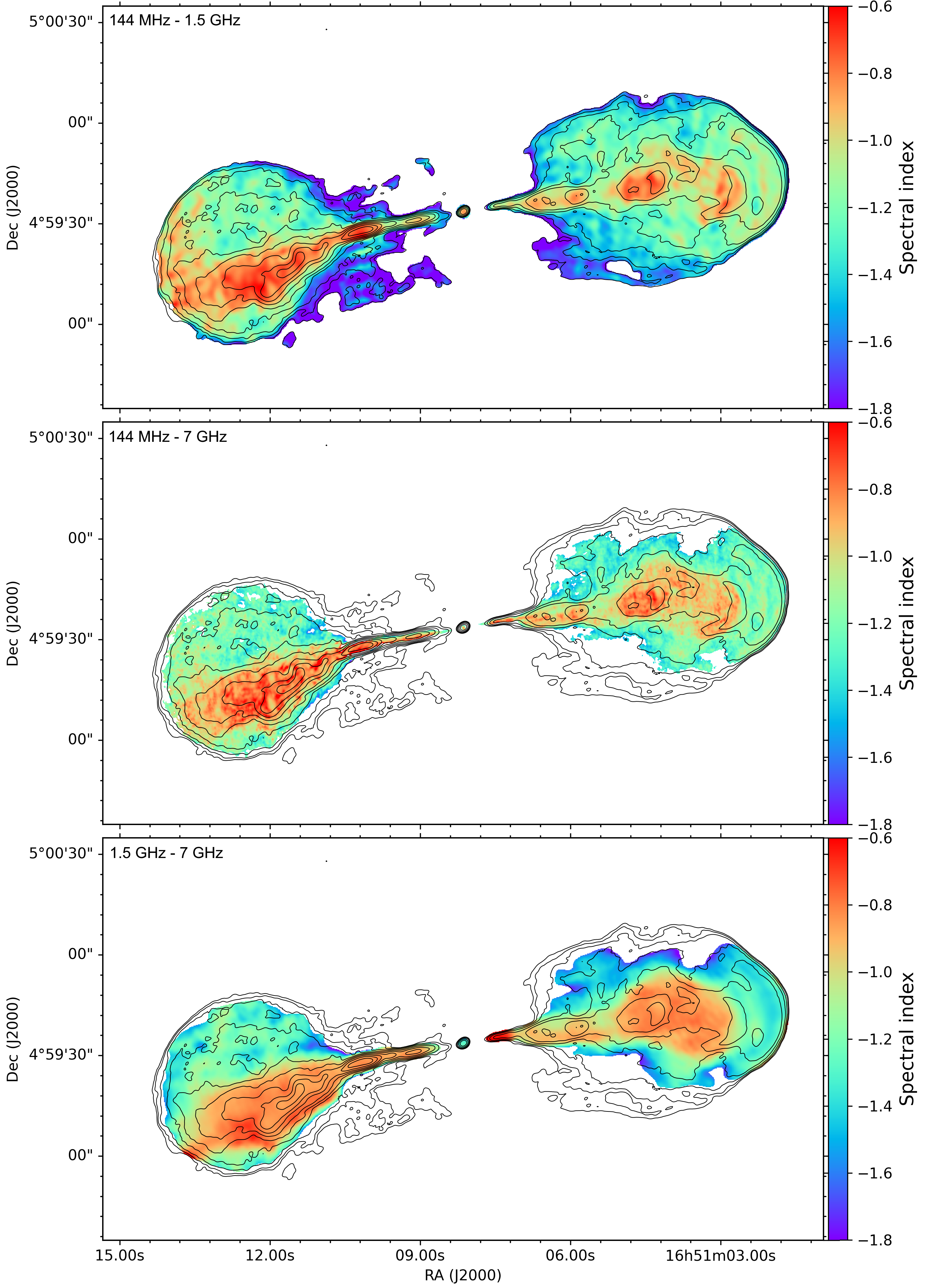}
    \caption{Spectral index maps of Hercules\,A between 144~MHz and 1.5~GHz (top panel), 144~MHz and 7~GHz (middle panel), and 1.5~GHz and 7~GHz (bottom panel). The contours indicate the L-band emission (1.5~GHz), and are drawn at [1, 2, 4, 8, ...] \(\times\ 5\sigma_\textrm{rms}\), where \(\sigma_\textrm{rms}=224.2\) µJy beam\(^{-1}\).}
    \label{fig:spixmaps}
\end{figure*}

\subsection{LOFAR}

Hercules\,A was observed with LOFAR's High Band Antennas \citep[HBA,][]{haarlem13} at frequencies between 120 and 168 MHz (PI: Timmerman, Project code: LC14-019). The data were recorded with spectral channels of 12~kHz to cover a total bandwidth of 48~MHz and with a time resolution of 1 second per integration. The observation took place on 8 June 2020, for a duration of 4~hours. The gain and bandpass calibrator source 3C\,295 was observed for 10~minutes before and after the target scan. The initial data reduction was performed using \textsc{Prefactor} \citep{vanweeren16, williams16, gasperin19}, which performed the initial flagging and used a model of the calibrator source to derive the calibration solutions for the Dutch stations. In particular, \textsc{Prefactor} first derived the polarization alignment and Faraday rotation. Based on these solutions, it derived the bandpass calibration solutions. Next, the clock corrections were derived by clock-total electron content (TEC) separation. These calibration solutions were then applied to the target data, after which the data were flagged again and averaged to a time resolution of 8 seconds and frequency channels of 98 kHz. Finally, a sky model of the target was obtained from the TIFR GMRT Sky Survey (TGSS) and used to perform a phase-only calibration cycle.

\begin{figure*}
    \centering
    \includegraphics[width=0.9\textwidth]{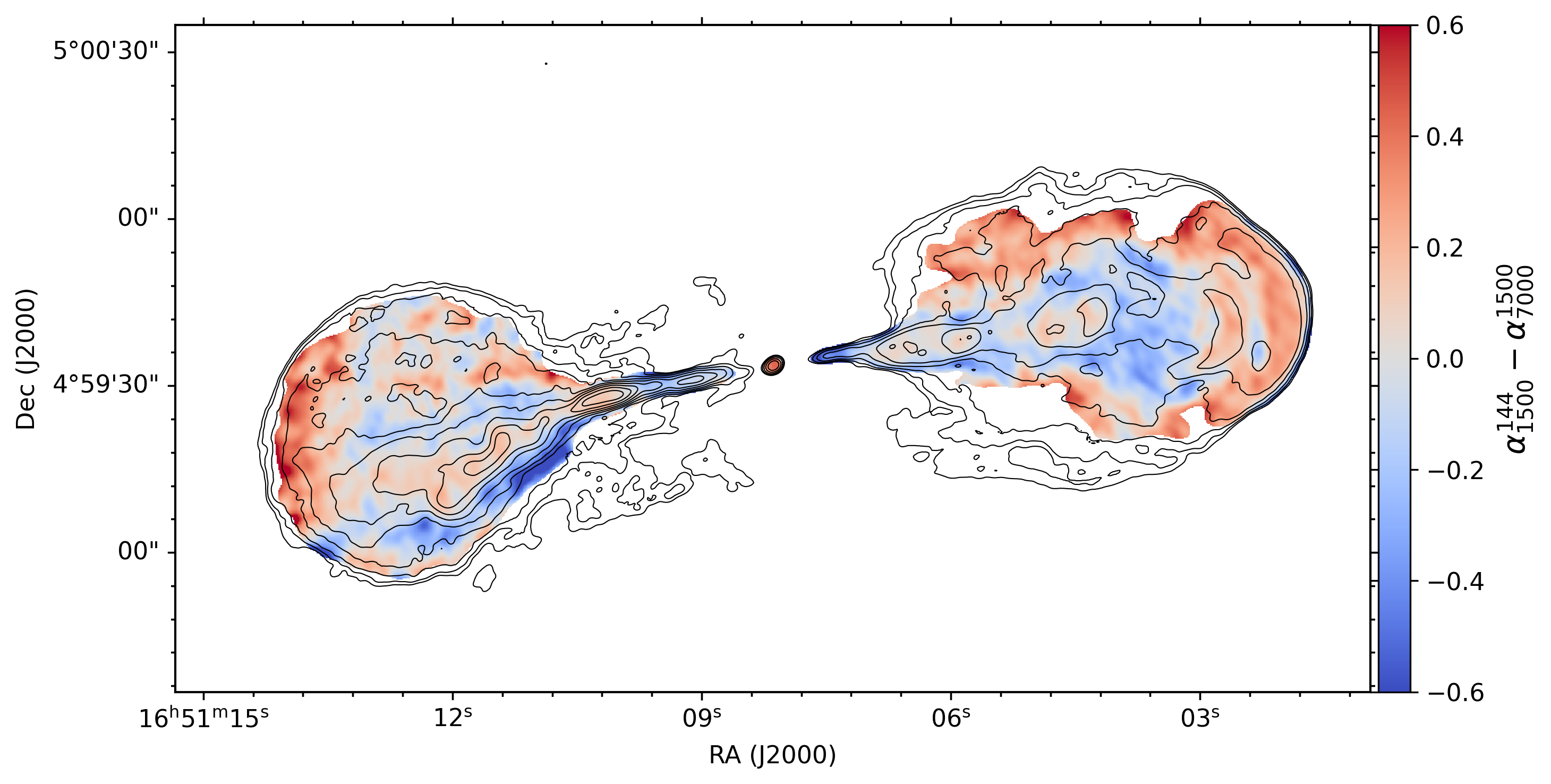}
    \caption{Spectral index curvature map of Hercules\,A. The spectral index curvature is calculated as the spectral index between 144~MHz and 1.5~GHz minus the spectral index between 1.5~GHz and 7~GHz. The red regions indicate where the spectrum is steeper at high frequencies than at low frequencies, whereas the blue regions indicate where the spectrum is flatter at high frequencies than at low frequencies. The contours indicate the L-band emission (1.5~GHz), and are drawn at [1, 2, 4, 8, ...] \(\times\ 5\sigma_\textrm{rms}\), where \(\sigma_\textrm{rms}=224.2\) µJy beam\(^{-1}\).}
    \label{fig:spix_curvature}
\end{figure*}

After this, the LOFAR-VLBI pipeline developed by \cite{morabito21} was employed to perform the initial calibration of the international LOFAR stations. This pipeline applied the \textsc{Prefactor} solutions to the unaveraged data containing all stations. Then, the pipeline identified the best in-field calibrator source from the Long Baseline Calibrator Survey \citep[LBCS,][in prep.]{jackson16}, and used this to derive the antenna delays by solving for the TEC. In the case of this observation, the target source Hercules A was used for this calibration step due to its high flux density. Finally, these calibration solutions were applied to the data to obtain the final calibrated data set.

To further improve the quality of the calibration, we performed phase and amplitude self-calibration on Hercules\,A using the Default Preprocessing Pipeline \citep[\textsc{DPPP},][]{vandiepen18} for deriving and applying updated calibration solutions and \textsc{wsclean} \citep{offringa14} for imaging. The final imaging was performed using the Briggs weighting scheme \citep{briggs95} with a robust parameter of -1, according to the Common Astronomy Software Application \citep[\textsc{CASA};][]{mcmullin07} definition. To calibrate the flux density scale, we scaled the final image (see Fig. \ref{fig:images}) to match the data from \citet{kellerman69}, where we interpolated their flux density measurements between the two nearest frequencies, which are 38 and 178~MHz. The low-frequency measurements of \citet{kellerman69} captured all of the flux density of Hercules\,A due to their low-resolution observations. A correction factor of 1.09 was applied to the data from \citet{kellerman69} to convert their measurements to the \citet{scaife12} flux density scale, which is identical to the RCB flux density scale \citep{roger73} at these low frequencies. We assume a 10\% uncertainty on the absolute flux scale due to intrinsic problems at low frequencies.

\subsection{Karl G. Jansky Very Large Array}

In addition to the LOFAR observations, Hercules\,A was observed with the VLA in the L band (1--2 GHz) and the C~band (6--8~GHz) between September 2010 and September 2011 (Project code: TDEM0011). The L-band observations were performed in the A, B, and C~configurations, while the C-band observations were also performed in D~configuration. The data were recorded in full polarization mode, with spectral channels of 1~MHz for the L-band observations and 2~MHz for the C-band observations. To exploit the complete 4~GHz of bandwidth available in the C~band, the observations were performed separately for the lower and upper halves of the C~band. The lower half of the C-band data were recently presented by \citet{wu20}, and the upper half of the C-band data are now presented along with the L-band observations for the first time. To calibrate both the flux density scale and the bandpass, scans on 3C\,286 were included in the observations. PKS\,J1651+0129 was used for phase-reference calibration. The flux density scale of the VLA observations was set to the \citet{perley2017} scale, which is consistent with the \citet{scaife12} flux density scale to within 5\%. 

The data reduction was performed using the Obit software package \citep{cotton08}. First, the data was Hanning smoothed and flagged. Next, the parallactic angle corrections, the delay calibration solutions, and the bandpass calibration solutions were determined. Using these solutions, the gains of the calibrator sources were derived based on source models and the flux density scale of the phase reference source was determined. Following this calibration procedure the data was flagged again, and the entire calibration was repeated with all flags applied from the start. Finally, the calibration solutions were applied to the data. To improve the calibration, Hercules\,A and the calibrator source, 3C\,286, were both self-calibrated. Based on the refined model of 3C\,286, corrections to the amplitude calibration were derived and applied to Hercules\,A. Hercules\,A was first self-calibrated based on the individual data sets. Then, these data sets were combined based on their spectral bands, and finally self-calibrated again. The imaging of the final data products was performed with multi-resolution CLEAN using the Briggs weighting scheme with robust parameters of -2.5 (L band) and -1 (C band), according to the Astronomical Image Processing System (AIPS) definition. We assume a 5\% uncertainty on the absolute flux scale in accordance with \citet{perley2017}.

\section{Results}

\begin{table}
    \caption{Properties of the images shown in Fig. \ref{fig:images}.}
    \centering
    \begin{tabular}{l|lll}\hline\hline
         & HBA & L band & C band\\\hline
        Observatory & LOFAR & VLA & VLA\\
        Frequency (MHz) & 144 & 1500 & 7000\\
        Bandwidth (MHz) & 48 & 1024 & 2048\\
        Total flux density (Jy) & 474.7 & 51.5 & 11.2\\
        Flux scale uncertainty & 10\% & 5\% & 5\%\\
        rms noise (µJy beam\(^{-1}\)) & 395.4 & 224.2 & 6.1\\
        b\(_\mathrm{major}\) (") & 0.547 & 2.085 & 0.396\\
        b\(_\mathrm{minor}\) (") & 0.234 & 1.592 & 0.349\\
        b\(_\mathrm{PA}\) (deg. East of North) & 1.602 & -59.930 & 86.550\\\hline\hline
    \end{tabular}
    \label{tab:img1}
\end{table}

\begin{figure*}
    \centering
    \includegraphics[width=0.7\textwidth]{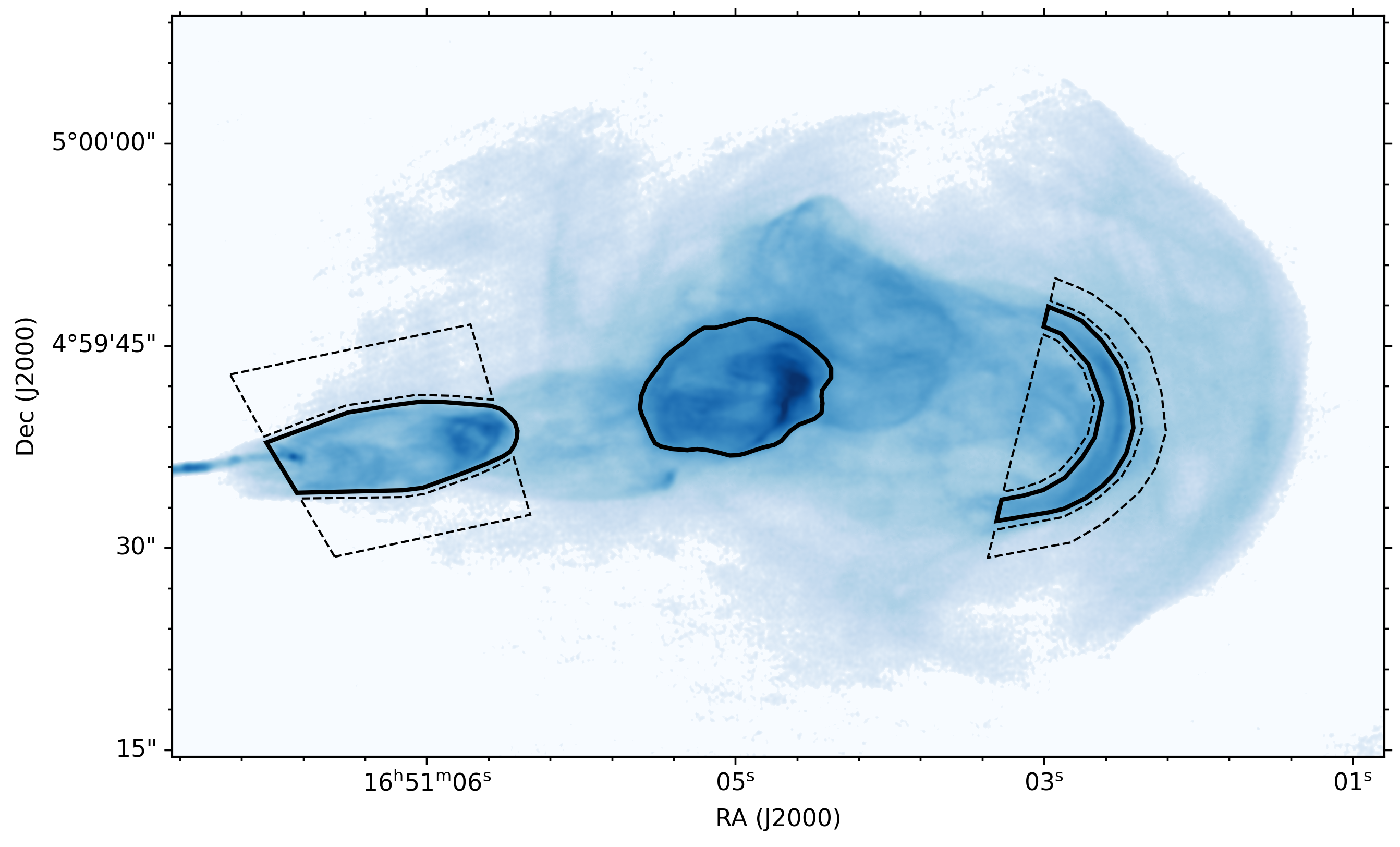}
    \caption{VLA image of the western lobe of Hercules\,A in C~band (7~GHz). The solid black regions indicate the regions used to measure the flux density of the rings. The regions with the dashed black borders indicate the regions used to estimate the background surface brightness of the lobe material at the position of the rings.}
    \label{fig:regions}
\end{figure*}

To investigate the nature of the rings, spectral index maps were produced between all three observing bands, as shown in Fig. \ref{fig:spixmaps}. To ensure a consistent sensitivity to the different spatial scales, upper and lower limits on the baseline length were applied to the LOFAR data to match both the L- and C-band VLA data. This reduced the largest angular scale to which the data are sensitive to 4.2~arcminutes in the case of a match with the C-band data and 20~arcminutes in the case of a match with the L-band data, both of which exceed the maximum angular scale of Hercules\,A (190~arcseconds). The resulting images were convolved to the same synthesized beam and confirmed to be aligned based on the central compact source. Finally, all regions below a 5\(\sigma\) confidence level in the spectral index map were masked out. Tests were performed to ensure that bowl effects due to the lack of short baselines do not introduce significant features in the spectral index maps.

In each of the spectral index maps, the jets are found to feature a spectral index of around \(\alpha=-0.8\) between the three bands, which is flatter than the typical values derived for the lobes. In the eastern lobe the flow of jetted plasma can be clearly traced by the gradual steepening of its spectrum as it ages and merges into the lobe. The jet appears to slowly diffuse, and to reach the outer edge of the lobe before diffusing into the lobe and eventually sinking back towards the central galaxy. In the western lobe the jetted plasma can mostly be seen as a relatively flat-spectrum (\(\alpha > -1\)) region in the center of the lobe. In addition, the rings are clearly distinguishable in the spectral index map between 144~MHz and 1.5~GHz. The spectral index of the bridge in between the two radio lobes remains poorly constrained, as this structure is too faint in the VLA maps, but will, in any case, have to be as steep as \(\alpha=-1.8\) or less to be consistent with a non-detection in the L-band image (1.5~GHz).

For use in the Discussion (Section 4), we calculate the difference between the spectral index maps between 144~MHz and 1.5~GHz, and 1.5~GHz and 7~GHz (Fig. \ref{fig:spix_curvature}). This difference allows us to determine whether the spectral index is constant from low to high frequencies, or if the spectral index changes. This can be interpreted as a measure of the curvature of the spectrum. In the spectral curvature map, the red regions indicate that the spectrum is steeper at high frequencies, whereas the blue regions indicate that the spectrum is steeper at low frequencies. In the western lobe, the three ring structures are distinguishable and feature steepening spectra. Interestingly, each ring shows more steepening than the previous one, counting from the core. The ring closest to the core shows an average spectral index curvature of \(0.02\pm0.07\). The ring in the middle shows an average spectral index curvature of \(0.04\pm0.07\). Finally, the farthest ring from the AGN shows an average spectral index curvature of \(0.16\pm0.07\). We note that the uncertainties on these measurements are dominated by systematic uncertainties, and are thus strongly spatially correlated.

\section{Discussion}

The origin of the ring structures in the western lobe has been one of the main topics of debate regarding Hercules\,A since their discovery by \citet{dreher84}. Currently, the shock model \citep[e.g.,][]{mason88, meier91} and inner-lobe model \citep[e.g.,][]{morrison96} remain in consideration, as neither model could be easily given preference \citep{gizani03}. However, our new observations extend the observed frequency range to much lower frequencies, while maintaining sufficient angular resolution to clearly resolve the rings. This allows us to measure the spectral properties of the rings over more than a decade of frequency, facilitating a comprehensive test of which model fits the data best.

The shock model provides one explanation for the morphology of the rings. In this model the rings are the shocks which induce adiabatic compression and particle acceleration when irregularities in the jet pass through the lobe. This scenario is able to recover the observed structure in the eastern lobe \citep{gizani03}. However, this model struggles to provide a compelling explanation for the observed spectral indices of the rings in the western lobe. First, the rings are found to have a relatively flat spectral index. Between our two lowest frequency bands we observe spectral indices of the rings in the western lobe of around \(\alpha=-0.8\), which is consistent with the spectral index of the eastern jet. The true value is most likely flatter than this as the observed spectral index includes contributions from the steeper-spectrum lobe material along the line of sight to the ring. Such flat spectral indices are unlikely to be produced merely through adiabatic compression, as adiabatic compression can only straighten the spectrum towards the injection spectral index \citep{ensslin01}, and even the low-frequency spectral index of the lobe is about \(\alpha=-1.2\). In comparison, the spectra of the rings at high frequencies are flatter than the spectrum of the lobes at low frequencies, indicating that particle acceleration is required as well. However, if particle acceleration by shocks plays a role, then it is difficult to explain why the outermost ring in the western jet features the steepest spectrum. The outermost ring forms the bow shock ahead of the western jet as it is caused by the frontal collision between the jet and the lobe. Therefore, it should be the strongest shock of the set. To resolve this, the outermost ring requires a significantly different Mach number and mixture between thermal and relativistic components compared to the other rings.

An alternative model posits that the rings in the western lobe are the surfaces of inner lobes created by the jet \citep{gizani03}. As the western jet recedes away from us, the brightness of this jet is strongly reduced by Doppler dimming. However, as the jets form inner lobes within the old plasma, the surface of this jetted plasma is decelerated, which reduces their Doppler dimming. This gives the surface of these inner lobes an apparent increase in brightness from our perspective. In the inner-lobe model, the bright rings are expected to be formed by the jetted material, and should therefore be similar to the jet in terms of spectral index, which is consistent with our observations. In particular, it is not problematic that the outer ring has the steepest spectrum of the set due to radiative aging since particle acceleration and adiabatic compression are not expected to contribute significantly. However, the inner-lobe model does have difficulty explaining the morphology of the rings. Inner lobes generally do not feature rim brightening, although this could at least partially be explained if the inner lobes are mainly filled with the Doppler-beamed jetted stream \citep{gizani03}. The inner-lobe model implies that the AGN experiences intermittent periods of activity, similar to that observed in the Perseus cluster \citep{fabian06}, Hydra\,A \citep{wise07}, MS0735 \citep{vantyghem14}, some peaked-spectrum radio sources \citep{Callingham2017}, and the Phoenix cluster \citep{timmerman21}, for instance. In the case of Hercules\,A evidence of such intermittency has already been provided by \citet{gizani02}, who detect a possible pair of parsec-scaled radio jets from a new outburst, and \citet{odea13}, who suggest based on Hubble Space Telescope observations that the AGN experienced an episode of activity about 60~Myr ago, and that it restarted about 20~Myr ago.

Accurate spectral index measurements can provide valuable information on the composition and physical conditions present in the emission region. Our low-frequency data allow us to not only accurately determine the overall spectral index, but also detect curvature in the spectrum. In the shock model the rings are formed by adiabatic compression and particle acceleration in the lobe. Both of these mechanisms work to straighten out the spectrum towards a single power-law model. On the other hand, the inner-lobe model predicts that the rings should feature the same spectral properties as the jet, which is subject to aging. In a synchrotron-emitting plasma, the higher-energy electrons radiate away their energy first, causing a break in the synchrotron spectrum \citep[e.g.,][]{scheuer68}. As the plasma ages, this break frequency shifts towards lower frequencies, leading to an increasingly steep spectrum at higher frequencies.

\begin{figure*}
    \centering
    \includegraphics[width=\textwidth]{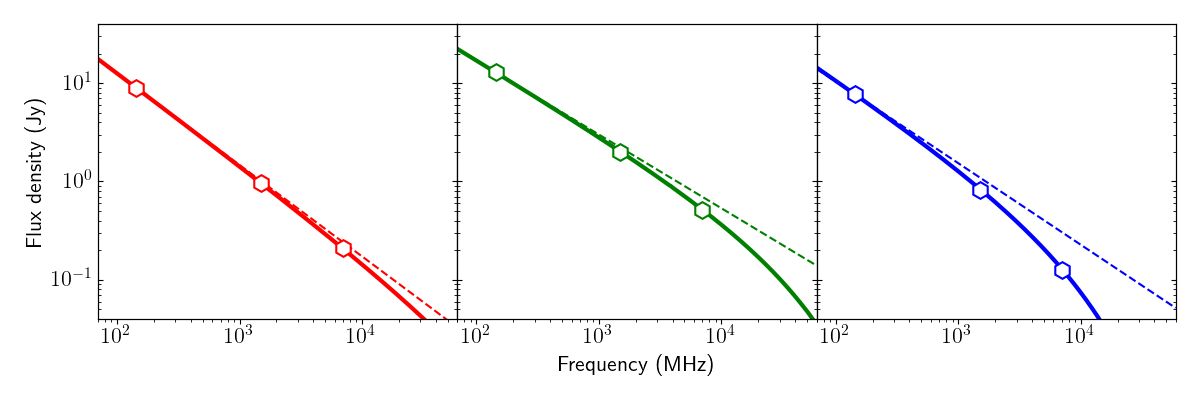}
    \caption{Illustration of the spectral curvature of the rings in the western lobe, assuming a Jaffe-Perola model \citep{jaffe73}. From left to right, the three panels show the inner, middle, and outer rings with injection indices of \(\alpha_\mathrm{inj}=-0.93\), \(\alpha_\mathrm{inj}=-0.75\), and \(\alpha_\mathrm{inj}=-0.83\), respectively. The hexagons indicate the total flux densities of the rings at the three different frequencies. The error bars on these measurements are drawn, but do not extend beyond the marker. The solid lines show the spectra of the three rings with the amount of curvature as measured from the flux density measurements without background subtraction. The dashed lines show the respective injection spectra without any curvature for comparison.}
    \label{fig:illustration}
\end{figure*}

As presented in Fig. \ref{fig:spix_curvature}, we observe a trend where each consecutive ring, counting from the core, shows more spectral steepening. To exclude the possibility that the spectral curvature trend emerges due to variations in the lobe emission along the line of sight, we also measured the spectral curvature by estimating the flux density contribution of the rings independent of the foreground material (see Fig. \ref{fig:images}). We did this by first measuring the total flux density within the ring region, and then subtracting the flux density due to the background surface brightness of the lobe. The regions used for this measurement in the C band are shown in Fig. \ref{fig:regions}. The background level is estimated as the mean surface brightness of two regions on opposite sides of the rings. We assume a 10\% uncertainty on all flux density measurements. Using the total flux density measurements without background subtraction, we obtained spectral curvature values \(\alpha_\mathrm{curv}=\alpha_{1500}^{144}-\alpha_{7000}^{1500}\) of \(0.04 \pm 0.13\), \(0.09 \pm 0.13\), and \(0.27 \pm 0.13\) for the inner, middle, and outer rings, respectively. For the background subtraction we excluded the middle ring from the analysis as the complex morphology of the ring and its environment prevent a reliable background level from being estimated. This leaves only the inner and outer rings, for which we estimate background-subtracted spectral curvatures of \(0.15\pm0.13\) and \(0.36\pm0.13\), respectively. 

We measure an increasing spectral curvature of the rings before and after background subtraction, implying that we are detecting the typical break in the synchrotron spectrum as a consequence of aging. This is also consistent with the similar spectral curvature values we found between the inner and middle ring, with the outer ring showing much stronger curvature. An illustration of the complete spectrum of each of the rings is shown in Fig. \ref{fig:illustration}. A Jaffe-Perola spectral model \citep{jaffe73} was fitted to the flux density measurements of each of the three rings to derive their respective injection index and break frequency. In the illustration, the break frequencies of the rings shift to lower frequencies the farther the rings are from the central AGN, causing the spectrum to steepen. We find relatively steep injection indices for the three rings of \(\alpha=-0.75\) up to \(\alpha=-0.93\). However, similar and steeper injection indices are not uncommon \citep[e.g.,][]{birzan08, harwood13, harwood15, shulevski15}. Although this model is strongly based on Doppler effects, it should be noted that the spectra will be significantly redshifted and will undergo beaming effects. However, as the rings all have a reduced Doppler factor due to the deceleration of the jetted plasma, the redshift of their spectra will be relatively low. This prevents the redshift of the rings from significantly affecting their relative break frequencies. 

To investigate whether the observed amount of spectral curvature is realistic for Hercules\,A, we derived the magnetic field strength in the jet required to produce such spectra. From our observations, we measure angular distances between the three rings of about 24 arcseconds, with a distance between the core of the AGN and the innermost ring of 37 arcseconds. This corresponds to projected physical distances of 66~kpc between the rings and 103~kpc between the core and the innermost ring. Assuming an angle between the jets and the line of sight of 50 degrees \citep{gizani99} and a jet speed of \(\beta=0.8\) \citep{gizani03}, we find that the innermost, middle, and outermost rings are composed of plasma with ages of approximately 550~kyr, 900~kyr, and 1250~kyr, respectively. Here, we assume that the inner lobes propagate at the speed of the jets, which is a rough approximation and is likely an upper limit to the true speed. However, more accurate measurements are unavailable, and the main goal is only to test whether the approximate numbers are sensible. Given these ages, and assuming a Jaffe-Perola spectral model, we derive magnetic field strengths in the jet of 28~µG, 36~µG, and 45~µG for the innermost, middle, and outermost rings, respectively. This trend in increasing magnetic field strength could be an indication of a deviation in the age estimates due to errors in the line-of-sight angle or jet speed, but it should be noted that there are also additional uncertainties due to the assumption of a Jaffe-Perola model, the methodology of measuring spectral curvature and the fact that there may be a small amount of redshift due to the recession of the jet. However, the approximate values of the magnetic field strength under the assumption of an inner-lobe model are in accordance with typical values derived for the jets of AGNs \citep[e.g.,][]{kataoka05, godfrey13}.

In addition to the spectral steepening of the rings, we also see that the environment of the jets shows spectral flattening at high frequencies, as shown in Fig. \ref{fig:spix_curvature}. This could be produced, for instance, by particle acceleration in regions that are opaque at low frequencies, or it could indicate a mismatch in the surface brightness sensitivities of LOFAR and the VLA. Such a discrepancy can emerge due to differences in the $uv$-coverage of the two arrays. LOFAR in particular suffers from a sparse $uv$-coverage at intermediate scales (1--2~arcsecond). However, the total flux density of the source has been carefully calibrated across all three frequencies and the consistent measurements obtained using different methods indicate the validity of our results.

\section{Conclusions}

In this paper we presented new LOFAR and VLA observations of Hercules\,A with the aim of investigating the nature of the ring features seen in the radio lobes. Our LOFAR observations clearly resolve the jet and ring structures in the lobes at 144~MHz, enabling the lobes to be studied at low frequencies, for the first time at a sub-arcsecond angular resolution. The bridge of emission in between the two lobes is only significantly detected at 144~MHz, which implies a spectral index of \(\alpha=-1.8\) or steeper to be consistent with a non-detection in our L-band observations (1.5~GHz).

To study the nature of the rings in the lobes of Hercules\,A, we investigate whether our observations are consistent with a shock model or an inner-lobe model. Spectral index mapping between 144~MHz and 7~GHz reveals that the rings in the western lobe feature spectral steepening at high frequencies. In particular, the spectral steepening increases as the rings are farther away from the central AGN. This suggests that the rings are subject to synchrotron aging, which is a clear prediction of the inner-lobe model. Therefore, we conclude that the observations presented in this paper are more consistent with the inner-lobe model where the jetted material from intermittent periods of AGN activity inflates small lobes within the outer lobe, which appear as ring-like structures. However, we do not exclude that adiabatic compression and particle acceleration contribute to the brightness of the rings as well, and we note that definitive evidence to settle this debate remains yet to be found.

Even though this model contributes evidence in favor of the inner-lobe model, it still struggles to provide a complete explanation for the morphology of the rings. Supporting evidence could likely be obtained through detailed magnetohydrodynamical (MHD) simulations. The jetted outflows can be simulated based on our current understanding of the environment of Hercules\,A. By replicating the observed ring-like structures, their exact nature can be confirmed. In particular, MHD modelling would have to focus on reproducing the observed morphology and spectral properties of the rings in the western lobe, and on the lack of ring structures in the eastern lobe.

\begin{acknowledgements}
      RT and RJvW acknowledge support from the ERC Starting Grant ClusterWeb 804208. JRC thanks the Nederlandse Organisatie voor Wetenschappelijk Onderzoek (NWO) for support via the Talent Programme Veni grant. CO and SB acknowledge support from the Natural Sciences and Engineering Research Council (NSERC) of Canada. AB acknowledges support from the VIDI research programme with project number 639.042.729, which is financed by the Netherlands Organisation for Scientific Research (NWO). JM acknowledges financial support from the State Agency for Research of the Spanish MCIU through the ``Center of Excellence Severo Ochoa'' award to the Instituto de Astrof\'isica de Andaluc\'ia (SEV-2017-0709) and from the grant RTI2018-096228-B-C31 (MICIU/FEDER, EU). This paper is based (in part) on data obtained with the International LOFAR Telescope (ILT) under project code LC14-019. LOFAR \citep{haarlem13} is the Low Frequency Array designed and constructed by ASTRON. It has observing, data processing, and data storage facilities in several countries, that are owned by various parties (each with their own funding sources), and that are collectively operated by the ILT foundation under a joint scientific policy. The ILT resources have benefitted from the following recent major funding sources: CNRS-INSU, Observatoire de Paris and Université d'Orléans, France; BMBF, MIWF-NRW, MPG, Germany; Science Foundation Ireland (SFI), Department of Business, Enterprise and Innovation (DBEI), Ireland; NWO, The Netherlands; The Science and Technology Facilities Council, UK; Ministry of Science and Higher Education, Poland. The National Radio Astronomy Observatory is a facility of the National Science Foundation operated under cooperative agreement by Associated Universities, Inc.
\end{acknowledgements}

\bibliographystyle{aa}
\bibliography{refs}

\end{document}